\documentclass[prl,twocolumn,superscriptaddress,floatfix,nobalancelastpage]{revtex4-1}

\usepackage{graphicx,bm,times}

\begin{document}

\title{A de Haas-van Alphen study of the Fermi surfaces \\of superconducting LiFeP and LiFeAs}

\author{C. Putzke}
\affiliation{H.H. Wills Physics Laboratory, University of Bristol, Tyndall Avenue, Bristol, BS8 1TL, UK.}
\author{A.I. Coldea} \email[corresponding author:]{amalia.coldea@physics.ox.ac.uk}
\affiliation{Clarendon Laboratory, Department of Physics, University of Oxford, Parks Road, Oxford OX1 3PU, U.K.}
\author{I. Guillam\'{o}n}
\affiliation{H.H. Wills Physics Laboratory, University of Bristol, Tyndall Avenue, Bristol, BS8 1TL, UK.}
\author{D. Vignolles}
\affiliation{Laboratoire National des Champs Magn\'{e}tiques Intenses (CNRS), Toulouse, France.}
\author{A. McCollam}
 \affiliation{High Field Magnet Laboratory, IMM, Radboud University Nijmegen, 6525 ED Nijmegen, The Netherlands.}
\author{D. LeBoeuf}
\affiliation{Laboratoire National des Champs Magn\'{e}tiques Intenses (CNRS), Toulouse, France.}
\author{M.D. Watson}
\affiliation{Clarendon Laboratory, Department of Physics, University of Oxford, Parks Road, Oxford OX1 3PU, U.K.}
\author{I.I. Mazin}
\affiliation{Code 6393, Naval Research Laboratory, Washington, DC 20375, USA.}
\author{S. Kasahara}
\affiliation{Research Center for Low Temperature and Materials Sciences, Kyoto University, Sakyo-ku, Kyoto 606-8501,
Japan.}
\author{T. Terashima}
\affiliation{Research Center for Low Temperature and Materials Sciences, Kyoto University, Sakyo-ku, Kyoto 606-8501,
Japan.}
\author{T. Shibauchi}
\affiliation{Department of Physics, Kyoto University, Sakyo-ku, Kyoto 606-8502, Japan.}
\author{Y. Matsuda}
\affiliation{Department of Physics, Kyoto University, Sakyo-ku, Kyoto 606-8502, Japan.}
\author{A. Carrington} \email[corresponding author:]{a.carrington@bristol.ac.uk}
\affiliation{H.H. Wills Physics Laboratory, University of Bristol, Tyndall Avenue, Bristol, BS8 1TL, UK.}

\begin{abstract}
We report a de Haas-van Alphen (dHvA) oscillation study of the 111 iron pnictide superconductors LiFeAs with $T_c
\approx 18$~K and LiFeP with $T_c \approx 5$~K. We find that for both compounds the Fermi surface topology is in good
agreement with density functional band-structure calculations and shows quasi-nested electron and hole bands. The
effective masses generally show significant enhancement, up to $\sim 3$ for LiFeP and $\sim$ 5 for LiFeAs. However, one
hole Fermi surface in LiFeP shows a very small enhancement, as compared with its other sheets. This difference probably
results from $\bm{k}$-dependent coupling to spin fluctuations and may be the origin of the different nodal and nodeless
superconducting gap structures in LiFeP and LiFeAs respectively.
\end{abstract}

\date{\today}
\maketitle

\affiliation{H.H. Wills Physics Laboratory, University of Bristol, Tyndall Avenue, Bristol, BS8 1TL, UK.}

\affiliation{Clarendon Laboratory, Department of Physics, University of Oxford, Parks Road, Oxford OX1 3PU, U.K.}

\affiliation{H.H. Wills Physics Laboratory, University of Bristol, Tyndall Avenue, Bristol, BS8 1TL, UK.}

\affiliation{Laboratoire National des Champs Magn\'{e}tiques Intenses (CNRS), Toulouse, France.}

\affiliation{Laboratoire National des Champs Magn\'{e}tiques Intenses (CNRS), Toulouse, France.}

\affiliation{Radboud University Nijmegen, HMFL, Faculty of Science,6500 GL Nijmegen, The Netherlands.}

\affiliation{Clarendon Laboratory, Department of Physics, University of Oxford, Parks Road, Oxford OX1 3PU, U.K.}

\affiliation{Code 6393, Naval Research Laboratory, Washington, DC 20375, USA.}

\affiliation{Research Center for Low Temperature and Materials Sciences, Kyoto University, Sakyo-ku, Kyoto 606-8501,
Japan.}

\affiliation{Research Center for Low Temperature and Materials Sciences, Kyoto University, Sakyo-ku, Kyoto 606-8501,
Japan.}

\affiliation{Department of Physics, Kyoto University, Sakyo-ku, Kyoto 606-8502, Japan.}

\affiliation{Department of Physics, Kyoto University, Sakyo-ku, Kyoto 606-8502, Japan.}

\affiliation{H.H. Wills Physics Laboratory, University of Bristol, Tyndall Avenue, Bristol, BS8 1TL, UK.}

Identification of the particular structural and electronic characteristics that drive superconductivity in the
iron-based materials continues to be a central experimental and theoretical question in the field. A successful theory
needs to explain trends, such as the variation of $T_{c}$ and also the structure of the superconducting energy gap. In
most of the iron arsenides the parent materials have a non-superconducting, antiferromagnetically ordered ground state.
Disruption of this magnetic order leads to superconductivity and then eventually a non-superconducting paramagnetic
ground state. A good example is the BaFe$_{2}$(As$_{1-x}$P$_{x}$)$_{2}$ series which has a maximum $T_{c}$ =30\,K when
$x\simeq 0.33$ \cite{Jiang09,Kasahara10}. Here BaFe$_{2}$As$_{2}$ has a magnetic ground state whereas BaFe$_{2}$P$_{2}$
is a paramagnet and neither superconduct.

\begin{figure}[tbp]
\center
\includegraphics*[width=8cm]{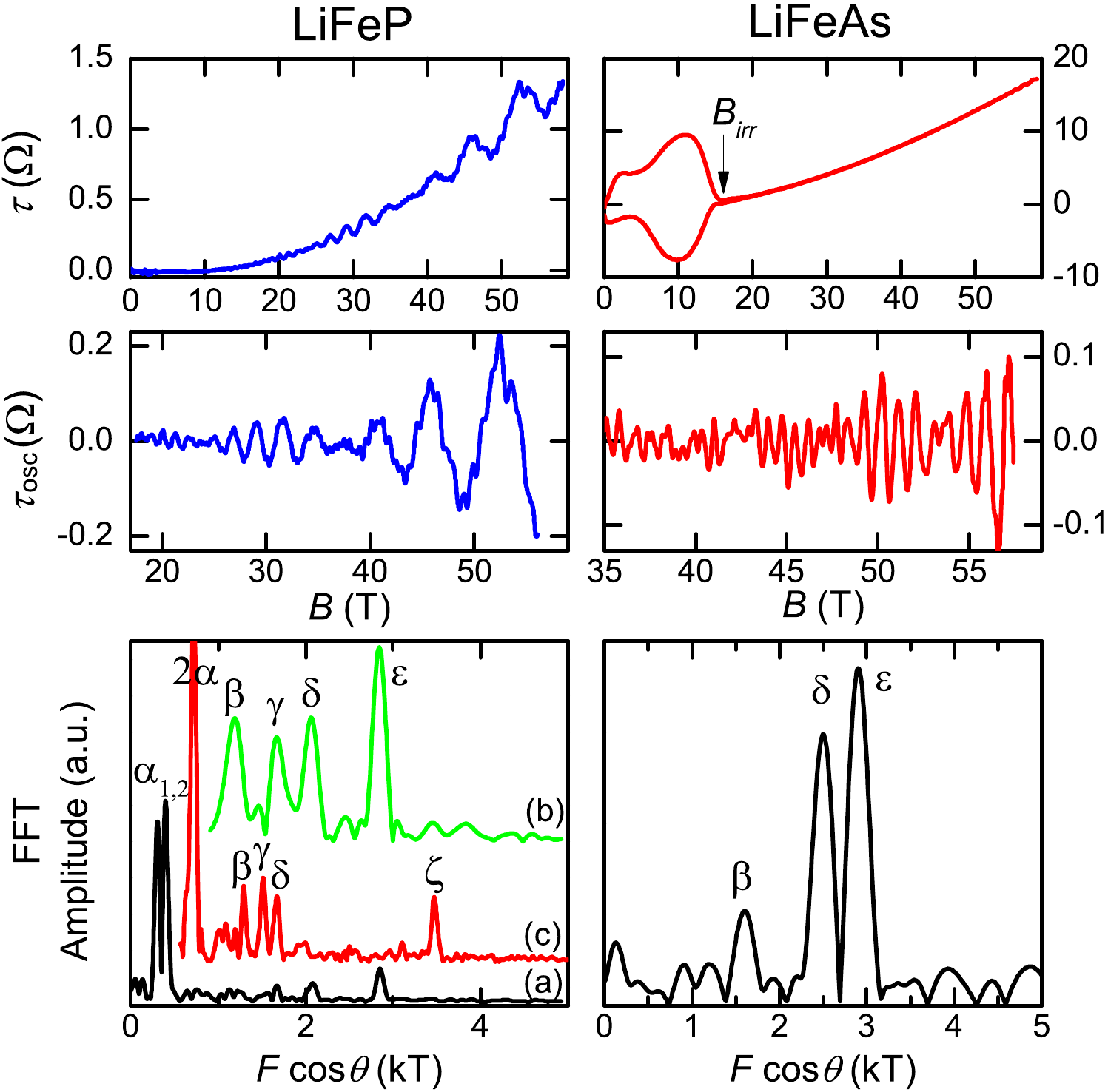}
\caption{(color online) Torque versus field for LiFeP and LiFeAs.
The top panels show the raw pulsed field torque data in units of
the change in the cantilever resistance at $T=1.5\,$K.
 The arrow
indicates the position of the irreversible field. The middle
panels show the oscillatory part of the torque after subtraction
of a smooth background. The bottom panels show FFTs of the torque.
For the peak labels see the main text. For LiFeP we show FFT
spectra computed over different field windows (a) (25-58\,T) which
shows the splitting of the $\alpha$ peaks, (b) (40-58\,T) which
decreases the influence of noise on the higher frequency peaks,
and (c) (33-45\,T) for the $dc$ field data at $T=0.45$\,K and
$\theta=51^\circ$, showing the strong $\zeta$ peak.}
\label{Fig:Raw}
\end{figure}

\begin{figure*}[tbp]
\center
\includegraphics*[width=18cm]{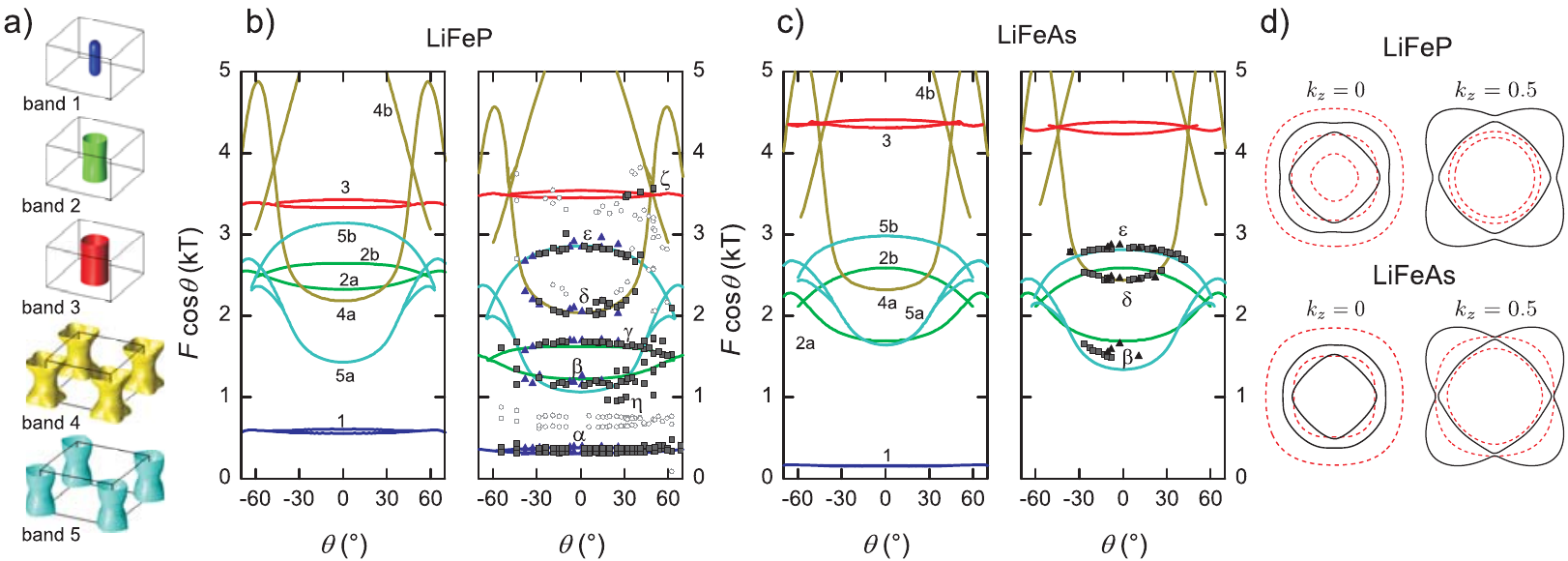}
\caption{(color online) (a) Calculated Fermi surfaces of LiFeP. (b) and (c) Show the evolution of de Haas-van Alphen
frequencies with magnetic field angle. Experimental data are shown in the right panels as symbols (triangle = pulsed
field, square/circle=dc field; circles=probably 2nd harmonics). The solid lines show the result of the DFT
calculations; the bands are shifted in the right hand panels to best fit the experimental results. The numbers refer to
the bands in (a). In all panels the frequencies have been multiplied by $\cos\protect\theta$ for clarity.(d) Slices
through the determined Fermi surfaces at particular $k_z$ values (with shifted bands). The dashed/solid lines are the
hole/electron sheets respectively, and the latter have been shifted along the [110] direction such that their center
coincides with the holes.} \label{Fig:Rotation}
\end{figure*}

The 111-family of iron-pnictides LiFeAs$_{1-x}$P$_{x}$, is unique because both LiFeAs and its counterpart LiFeP
superconduct and are non-magnetic with $T_{c}\sim 18$\,K \cite{Tapp08,Pitcher08} and $\sim 5$\,K \cite{Deng09},
respectively. Also, penetration depth measurements have shown that LiFeAs is fully gapped \cite{Kim11,Hashimoto11},
whereas LiFeP has gap nodes \cite{Hashimoto11}. Establishing whether this switch of pairing structure is linked to
changes in the topology and orbital character of the Fermi surface (FS) provides an stringent test of candidate
theories for the superconducting pairing in these materials.

Magneto-quantum oscillation effects such as the de Haas-van Alphen (dHvA) effect are a powerful probe of the
three-dimensional bulk Fermi surface and have been successfully used to study a variety of iron-based superconductors
\cite{Coldea10,Carrington11}. In this Letter, we present a study of the dHvA oscillations in both LiFeP and LiFeAs
which establishes that the \emph{bulk} Fermi surface topology of these compounds is in good agreement with DFT
calculations. Furthermore, by comparing the values of the extracted effective masses of the quasiparticles to the
calculated band masses, we find significant orbit dependence to the mass enhancement factors which is likely linked to
the contrasting superconducting gap structures and $T_c$ in these compounds.

Single crystals of LiFeP and LiFeAs were grown by a flux method \cite{Kasahara11}. Small single crystals, typically
$50\times 50\times 10\,\mu $m$^{3}$ for LiFeP and $100\times 200\times 50\,\mu $m$^{3}$ for LiFeAs, were selected for
the torque measurements. To avoid reaction with air the samples were encapsulated in degassed Apiezon-N grease. Sharp
superconducting transitions were measured using radio frequency susceptibility with $T_{c}$ onset (midpoint) values of
4.9\, K (4.7\, K) and 18.4\, K (17.3\, K) for LiFeP and LiFeAs, respectively. The samples were mounted onto miniature
Seiko piezo-resistive cantilevers which were installed on a rotating platform, immersed in liquid $^{4}$He, in the bore
of a pulsed magnet up to 58~T in Toulouse. Measurements on the same crystals were also conducted in an 18\,T
superconducting magnet in Bristol and a 33\,T Bitter magnet at HMFL in Nijmegen and 45\,T hybrid magnet at NHMFL,
Tallahassee, all equipped with $^{3}$He refrigerators.

Torque versus magnetic field data are shown in Fig.\ \ref{Fig:Raw}. For both materials dHvA oscillations are seen at
high fields and low temperatures, well above the upper critical field, estimated to be $\lesssim 1$\,T for LiFeP
\cite{Mydeen10} and $\sim$16\,T for LiFeAs \cite{Kurita11} when $B\|c$ (see also Fig.\ \ref{Fig:Raw}). After the fast
Fourier transform (FFT) as a function of inverse field, several strong peaks are visible (Fig.\ \ref {Fig:Raw} bottom
panels), which correspond to the extremal cross-sectional areas $A_{k}$ of the FS: $F=\hbar A_{k}/2\pi e$. For LiFeP,
the spectrum is dominated by two low dHvA frequencies around 300\, T and 400\, T, labelled $\alpha _{1}$ and $\alpha
_{2}$. The amplitude and frequency of the peak at $\sim 750$\, T is consistent with this being the second harmonic of
the $\alpha $ peaks. The other five peaks ($\beta, \gamma, \delta, \varepsilon, \zeta$) are clearly derived from unique
Fermi surface orbits. For LiFeAs, three frequencies are visible at 1.5\,kT, 2.4\,kT and 2.8\,kT, labelled as $\beta,
\delta, \varepsilon$, respectively.

To properly identify these FS orbits, we performed field sweeps with different field orientations starting from $\theta
=0^{\circ }$ ($B\Vert c$) and rotating towards the $ab$-plane. For a perfectly two dimensional (2D) FS, $F\propto
1/\cos \theta $, so by multiplying $F$ by $\cos \theta $ the degree of two dimensionality of a FS can be easily seen.
For quasi-2D surfaces, $F\cos \theta $ will decrease with increasing $\theta$ for a local maximum of Fermi surface
orbit area as a function of $k_{z}$ whereas the opposite will be true for a local minimum. The data in Fig.\
\ref{Fig:Rotation} suggest that for LiFeP $\varepsilon $ and $\gamma$ are a maxima, and $\beta $ and $\delta $ are
minima. The two lowest frequency $\alpha $ orbits have opposite curvature indicating that they are the maximum and
minimum of the same FS sheet.  At angles close to $\theta=50^\circ$ strong peaks are seen (labelled $\zeta$ in Figs.\
\ref{Fig:Raw} and \ref{Fig:Rotation}) which are likely from the outer hole sheet (band 3). The amplitude becomes large
at this angle because of the Yamaji effect, expected when the two extremal orbits of a quasi two dimensional Fermi
surface cross.  Close to $\theta=30^\circ$, an additional branch $\eta$ is visible. For LiFeAs, the $\varepsilon $
orbit is a maximum, while $\delta$ and $\beta$ orbits are likely to be minima orbits.

To identify the origin of the orbits and  solve the structure of the Fermi surface we have performed DFT calculations
using the linear augmented plane wave method, implemented in the \textsc{wien2k} package \cite{wien2k}. We used the
experimental crystal structure \cite{latticeparams} and included spin-orbit coupling (SOC).  The calculated Fermi
surfaces (see Fig.\ \ref{Fig:Rotation}(a,d)) are quite similar for both materials, there are three hole bands at
$\Gamma$ and two electron bands at M as found previously \cite{Shein10}. The two outermost hole sheets are quite 2D,
whereas the innermost $xz/yz$ hole pocket is strongly hybridized with $d_{z^{2}}$ near Z and is closed there, while
remaining 2D away from this point. By contrast, the electron orbits are very strongly warped. This geometry is
reflected in the calculated angular dependence of the dHvA orbits (Fig.\ \ref{Fig:Rotation}(b,c)). For the 2D hole
sheets $F\cos \theta $ varies little with angle and the maximal and minimal area are close. For the electron sheets
there is a large deviation from this behavior. For LiFeP, SOC splits the two outermost hole bands, which are
accidentally nearly degenerate in non-relativistic calculations, and causes their character to be mixed
$d_{xz/yz}/d_{xy}$. In LiFeAs these bands are well separated irrespective of SOC and have a predominantly $d_{xz/yz}$
(middle) and $d_{xy}$ (outermost) character. The SOC also splits the electron bands along the zone edge (X-M) inducing
a gap of $\sim$35\,meV (see \ref{Fig:Rotation}(d)), hence as in LaFePO \cite{Carrington09} we estimate that magnetic
breakdown orbits, along the elliptic electron surfaces in the unfolded Brillouin zone, to be strongly damped.

By comparing the calculations to the data (Fig.\ \ref{Fig:Rotation}(b,c)), in particular the curvature of $F\cos
\theta$, the correspondence between the observed dHvA frequencies and the predicted Fermi surface orbits is immediately
apparent for most orbits. The observed $\beta$ frequencies are likely a mixture of signals from orbits 2a (hole) and 5a
(electron) close to $\theta=0^\circ$ but are separately resolved at angles close to $30^\circ$ (the $\eta$ branch
probably corresponds to band 5a). For LiFeP, relatively small shifts (somewhat smaller than for other Fe pnictides
\cite{Carrington11}) of the band energies: $+20$\, meV and $+45$ meV for band 4 and 5 (electron) and $-65$, $-80$,
$18$\, meV for bands 1, 2 and 3 (hole) bring the observations and calculations into almost perfect agreement as shown
in Fig.\ \ref{Fig:Rotation}(b). As in other Fe pnictides \cite{Shishido10,Carrington11}, these shifts shrink both the
electron and hole FSs and likely originate from many body corrections to the DFT bandstructure \cite{Ortenzi09}.
Although the maximal orbit of band 4 which is close to 6\,kT was not observed, probably because the scattering rate in
our sample was too high, we can estimate the accuracy of our band energy determinations by calculating the difference
in total volume of the electron and hole Fermi surfaces. We find a small imbalance of just $+0.02$ electrons per unit
cell which shows the consistency of the procedure.

For LiFeAs, the curvature and absolute values of F$\cos \theta $ suggest that the $\varepsilon $ orbit originates from
the maximum of the inner electron Fermi surface (band 5) and the extended angular dependence of the $\delta$ orbit
suggest that this originates from the minima of the electron surface (band 4a), rather than the maximum of the middle
hole surface (band 2b) which is of similar size in the calculation. The limited angular extent of the data for the
$\beta$ orbit means it is not possible to say if it originates from band 5a (electron) or band 2a (hole) although 5a
seems more likely. To accurately match the $\varepsilon$ and $\delta$ orbits with the calculations only very small
shifts of the band energies are required ($-$5\,meV and $+18$\,meV for bands 4 and 5 respectively) (Fig.\
\ref{Fig:Rotation}(b)). We did not observe the smallest hole  FS (band 1)in LiFeAs, even though the same band gave the
largest signals for LiFeP. This suggests that band 1 does not cross the Fermi level in LiFeAs, which requires that it
shifts down by $\sim -40$\,meV, possibly because of enhanced SOC. The small shift of the electron bands is almost
perfectly compensated by the removal of band 1, so the remaining hole bands are not shifted in Fig.\
\ref{Fig:Rotation}(b). Although we do not see definitive evidence for the hole orbits, probably because of a
significantly higher impurity scattering rate in LiFeAs compared to LiFeP, the small size of the energy shifts needed
to match the electron bands combined with the similar small shifts required in LiFeP to match \textit{both} electron
and hole bands strongly suggests that the DFT calculations correctly predict the Fermi surface topology of these 111
compounds.  This is contrast to the photoemission results of Borisenko \emph{et al.} \cite{Borisenko10} for LiFeAs,
where a significant discrepancy between the size of the hole sheets and the DFT calculations was found.

\begin{figure}[tbp]
\center
\includegraphics*[width=8cm]{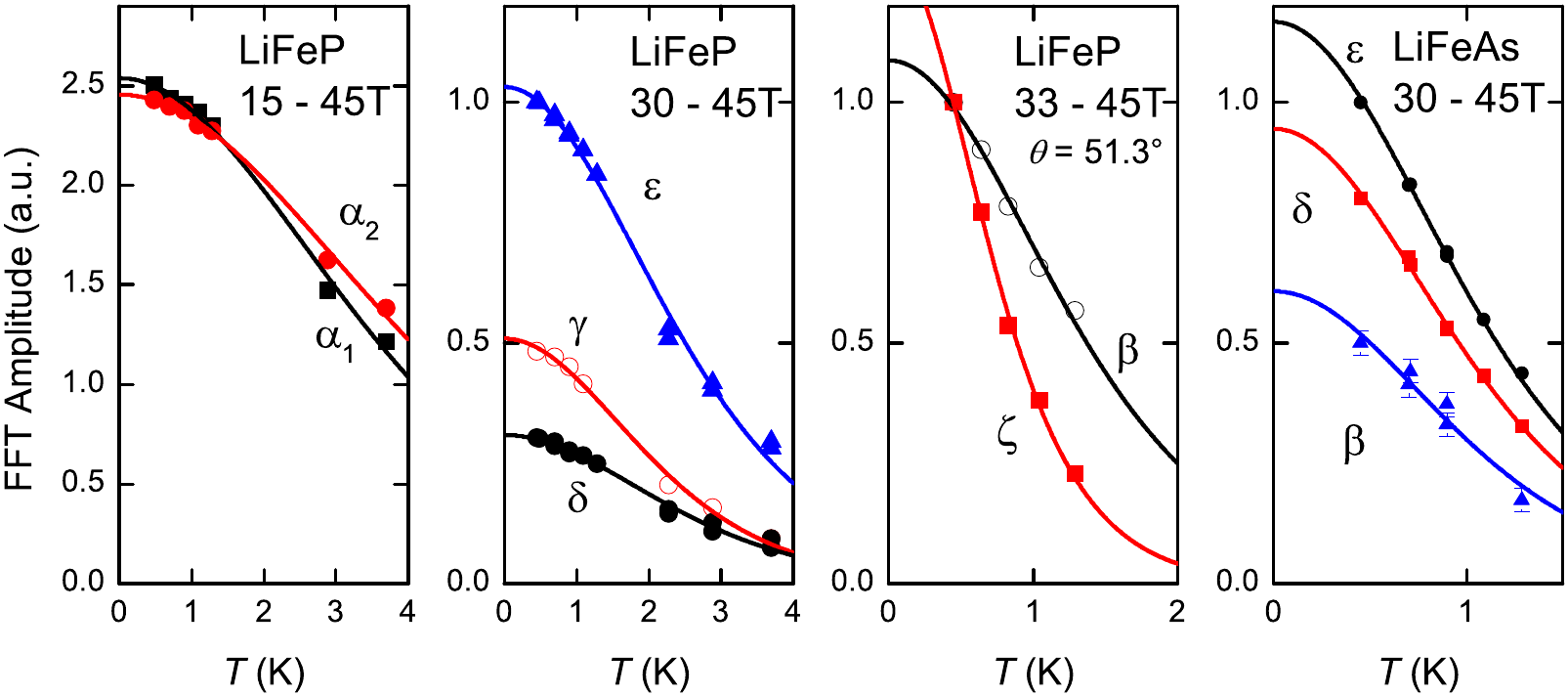}
\caption{(color online) Quasiparticle effective masses determination. Amplitude of the FFT peaks (the field ranges as indicated)
versus temperature. The lines are fits to the Lifshitz-Kosevich formula \protect\cite{Shoenberg}. The effective mass values are shown in
Table \ref{Table:mass}.}
\label{Fig:Mass}
\end{figure}

The strength of the electron-electron interactions can be estimated from measurements of the quasiparticle effective
mass $m^{\ast }$ on each orbit through the temperature dependence of the amplitude of the dHvA signals, by fitting the
latter to the Lifshitz-Kosevich formula \cite{Shoenberg} (Fig.\ \ref{Fig:Mass}). These measurements were conducted in
dc field on the same samples to avoid any possibility of sample heating at low temperature.  The derived values along
with the DFT calculations are shown in Table \ref{Table:mass}.

\begin{table}
\caption{Measured and calculated dHvA frequencies. The measured frequencies are extrapolated to $\theta$ = 0. The
effective ($m^*$) and calculated ($m_b$) band masses are quoted in units of the free electron mass ($-$ sign indicates
hole orbit).  For the LiFeP the values marked with  $\dagger$  were determined at an angle of $51\,^\circ$ ($\lambda$
was calculated with $m_b$ also calculated at this angle). At this angle orbits 3a and 3b cross (and have maximum
amplitude), and orbit 2a is clearly differentiated from 5a. Values marked with $\sharp$ potentially have overlapping
orbits and so the mass enhancements and orbit assignments are less certain.} \label{Table:mass}
\begin{tabular}{lcl@{\hspace{1cm}}lcrr}\hline
\multicolumn{7}{l}{LiFeP}\\
\multicolumn{3}{l}{DFT calc.}&\multicolumn{4}{l}{Experiment}\\
\hline
Orbit & $F(T)$ & $m_b$ & Orbit & $F(T)$ & $m^*$ & $\frac{m^*}{m_b}-1$\\
$1_a$   & 557    &$-0.44$  & $\alpha_1$& 316(2) & 1.1(1) & 1.5(3)\\
$1_b$   & 607    &$-0.39$  & $\alpha_2$& 380(2) & 1.0(1) & 1.6(3)\\
$2_a$   & 2325   &$-1.7$   & $\beta^\dagger$ & 2040(10)$^\dagger$ & 4.4(1)$^\dagger$ & 0.6(2)$^\dagger$\\
$2_b$   & 2645   &$-1.6$   & $\gamma$  & 1670(10) & 2.7(2) & 0.7(1)\\
$3_a$   & 3328   &$-1.8$   & $\zeta^\dagger$     & 5550(10)$^\dagger$  & 7.7(2)$^\dagger$& 2.1(5)$^\dagger$\\
$3_b$   & 3428   &$-1.6$   & $\zeta^\dagger$     & 5550(10)$^\dagger$  & 7.7(2)$^\dagger$& 2.1(5)$^\dagger$\\ \\
$4_a$   & 2183   &$+0.92$   & $\delta$  & 2040(20) & 2.2(1)& 1.4(2)\\
$4_b$   & 6014   &$+1.8$   & & & &\\
$5_a$   & 1430   &$+1.1$    &$\beta$$\sharp$   & 1160(10) & 3.6(2)$\sharp$& 2.3(2)$\sharp$ \\
$5_b$   & 3142   &$+0.83$   &$\varepsilon$ & 2840(10)& 2.2(2)& 1.6(3)\\
\hline
\multicolumn{7}{l}{LiFeAs}\\
\multicolumn{3}{l}{DFT calc.}&\multicolumn{4}{l}{Experiment}\\
\hline
Orbit & $F(T)$ & $m_b$ & Orbit & $F(T)$ & $m^*$ & $\frac{m^*}{m_b}-1$\\
$1_a$   & 130    &$-0.31$  \\
$1_b$   & 149    &$-0.23$  \\
$2_a$   & 1585   &$-2.11$  \\
$2_b$   & 2529   &$-1.50$  \\
$3_a$   & 4402   &$-2.11$  \\
$3_b$   & 4550   &$-2.12$  \\
$4_a$   & 2359   &$+1.22$  &$\delta$&2400(25)&5.2(4)& 3.3(3)\\
$4_b$   & 6237   &$+2.34$   \\
$5_a$   & 1584   &$+1.54$   &$\beta$$\sharp$&1590(10)& 6.0(4)$\sharp$ & 2.9(3)$\sharp$\\
$5_b$   & 2942   &$+1.02$ &$\varepsilon$&2800(40)&5.2(4) & 4.1(4) \\
\hline
\end{tabular}
\label{Table:massList}
\end{table}

For LiFeP, the enhancements factors $\lambda=m^{\ast }/m_{b}-1$ vary strongly between orbits.  For the electron sheet
 $\lambda$ is in the range 1.4-2.3, which is comparable to values found for the electron sheets of LaFePO
($T_{c}$=6\thinspace K) \cite{Coldea08}.  The smallest and largest hole sheets (bands 1 and 3) are also strongly
enhanced, however for the middle hole sheet (orbits $\gamma$ and $\beta$, band 2) $\lambda$ is $\sim 3$ times smaller
than for the other sheets, despite having similar orbital character. As an enhancement $\lambda\simeq 0.2$
\cite{Boeri09} is expected from electron-phonon coupling, this means that the residual electron-electron component for
this particular orbit is very weak. This is an interesting observation, relevant to the ongoing discussion
\cite{Hirshfeld11} as to whether the mass enhancement comes entirely from local correlations or partially from long
range spin fluctuations. If the mass renormalization in this compound is due to the same spin fluctuations that are
believed to cause superconductivity, we can conclude that band 2 is very weakly coupled with these fluctuations, so
that the pairing amplitude on this band will be small and hence it is a possible candidate for the location of the gap
nodes. Calculations suggest \cite{Hirshfeld11} that node formation is controlled by the $xy$ pocket, so that if this
pocket exists, the order parameter is nodeless, otherwise nodes form on an \textit{electron }(band 4, in our notation)
pocket. LiFeP seems to deviate from this rule, as it has a well developed $xy$ pocket (band 3) and has gap nodes. LiFeP
therefore appears to be a challenging and an extremely interesting material for further theoretical modelling.

For LiFeAs, the measured effective masses are uniformly larger than in LiFeP. For the electron sheet (band 5) $\lambda$
 is more than 3 times larger than in LiFeP.  This observation suggest that mass renormalization in iron pnictides
is caused \textit{by the same interaction that drives superconductivity}. This agrees with previous findings in the
isoelectronic superconducting series, BaFe$_2$(As$_{1-x}$P$_{x}$)$_2$, in which the effective mass of the electron
bands are closely related to the increase in $T_c$ \cite{Shishido10}. Interestingly, the large mass enhancement in
LiFeAs is not accompanied by a corresponding large shrinking of the Fermi surface volume \cite{Shishido10}.

In summary, dHvA oscillations have been observed in two members of the 111 family of superconductors, LiFeP and LiFeAs.
In both cases we find that measured data are consistent with the topology of the DFT calculated Fermi surface with
small band energy shifts. The many-body mass enhancements are larger in LiFeAs than in LiFeP. In LiFeP, the middle hole
band has significantly lower mass enhancement than the other sheets, which implies that the electron-hole scatter rate
is suppress for this sheet. This may be the origin of the lower $T_{c}$ and nodal gap in LiFeP, and suggests that the
mass enhancement is to a large extent due to a $\bm{k}$-dependent spin-fluctuation induced interaction, which are also
responsible for the pairing. It will be very interesting to see whether these features and the nodal gap structure in
LiFeP can be explained by detailed microscopic calculations.

We thank E. Kampert, F. Wolff-Fabris, E.A.\ Yelland and F.\ Fabrizi, E.\ Choi and A.\ Bangura for technical assistance
and S.\ Borisenko for discussions. This work is supported by EPSRC (UK), EuroMagNET II under the EU contract no.\
228043,  KAKENHI from JSPS. A portion of this work was performed at the National High Magnetic Field Laboratory, which
is supported by National Science Foundation Cooperative Agreement No. DMR-0654118, the State of Florida, and the U.S.
Department of Energy.

\bibliography{Li111dHvA}

\begin{thebibliography}{23}
\bibitem[{lat(????)}]{latticeparams}
\bibinfo{note}{The lattice parameters for LiFeP were determined by x-ray
  diffraction at 300\,K, $a$=3.6955\,\AA\ $c$=6.041\,\AA\ $z_{\rm Li}$=0.1437,
  $z_{\rm{P}}$=0.2803 and for LiFeAs we used: a=3.7914\,\AA\ c=6.3639\,\AA\
  $z_{\rm Li}$=0.1541, $z_{\rm{As}}$=0.2635 from Ref\ \cite{Tapp08}.}
\end{thebibliography}

\bibliographystyle{aps5etal}

\end{document}